# Vector dark domain wall solitons in a fiber ring laser

**H. Zhang, D. Y. Tang\*, L. M. Zhao and R. J. Knize**

[1]*School of Electrical and Electronic Engineering, Nanyang Technological University,*

*Singapore 639798*

[*]*Corresponding author: edytang@ntu.edu.sg*

We observe a novel type of vector dark soliton in a fiber ring laser. The vector dark soliton consists of stable localized structures separating the two orthogonal linear polarization eigenstates of the laser emission and is visible only when the total laser emission is measured. Moreover, polarization domain splitting and moving polarization domain walls (PDWs) were also experimentally observed.

OCIS *codes*: 060.4370, 060.5530, 140.3510.

Soliton operation of fiber lasers have been observed with anomalous cavity dispersion. It was found that the solitons formed could be described by the nonlinear Schrödinger equation (NLSE) [1]. Formation of solitons in a fiber laser is a result of the mutual nonlinear interaction among the laser gain and losses, cavity dispersion and fiber nonlinearity, as well as the cavity effects. Dynamics of the laser solitons should be described by the complex Ginzburg-Landau equation (CGLE) [2]. However, it was noticed that solitons formed in the anomalous dispersion cavity fiber lasers have normally a narrow spectral bandwidth, which is far narrower than the laser gain bandwidth. Consequently, no gain bandwidth filtering effect practically exists in the lasers, and the effect of laser gain is mainly to balance the cavity losses. It was confirmed experimentally when the effect of spectral filtering on the pulse shaping could no longer be ignored, the solitons formed in a fiber laser could not be described by the NLSE but the CGLE [3]. Solitons governed by the CGLE are also called the dissipative solitons. Recently, formation of dissipative solitons in fiber lasers has attracted considerable attention [4].

In addition to gain, the vector nature of light also needs to be considered for fiber lasers whose cavity consists of no polarizing components. In these fiber lasers the soliton dynamics is governed by the coupled NLSEs or CGLEs. Theoretical studies on the coupled NLSEs have shown that due to the cross polarization coupling, new types of solitons, such as the group velocity locked vector soliton [5], phase locked vector soliton [6], induced solitons [7], and high order phase locked vector solitons [8], could be formed. Indeed, these solitons were experimentally observed in fiber lasers. Moreover, PDWs and a novel type of vector dark PDW soliton were also theoretically predicted using coupled NLSEs [9]. However, they have not been observed in fiber lasers. Phenomena related to PDW soliton have been observed in passive fibers using two counter-propagating laser beams [10] and by polarization modulation instability [11]. In this letter we report the first experimental observation of the vector dark PDW solitons in a fiber ring laser.

Our experiment was conducted on a fiber laser schematically shown in Fig. 1. The ring cavity is made of all-anomalous dispersion fiber, consisting of 6.4 m, 2880 ppm Erbium-doped fiber (EDF) with group velocity dispersion (GVD) of 10 (ps/nm)/km and 5.0 m single mode fiber (SMF) with GVD of 18 (ps/nm)/km. A polarization insensitive isolator was employed in the cavity to force the unidirectional operation of the ring, and an in-line polarization controller (PC) was used to fine-tune the linear cavity birefringence. A 10% fiber coupler was used to output the signal. The laser was pumped by a high power Fiber Raman Laser source of wavelength 1480 nm. An in-line polarization beam splitter (PBS) was used to separate the two orthogonal polarizations of the laser emission, and they were simultaneously measured with two identical 2GHz photo-detectors and monitored in a multi-channeled oscilloscope.

A similar configuration fiber laser was previously investigated and a fast antiphase square-pulse polarization dynamics was observed [12]. The observed fast antiphase polarization dynamics was interpreted as caused by the gain competition between the two cavity polarization modes and the cavity feedback. Feedback induced polarization switching in fiber laser was also previously reported [13]. Antiphase square pulse emission along the two orthogonal polarizations was observed in our laser. However, it was different from previous observations [12] in that the square pulse width varied with the cavity birefringence. Fig. 2 shows an experimentally measured square-pulse width variation with the orientation of one of the PC paddles. In a range of the paddle's orientation the laser emitted square pulses, and the square pulse width could be continuously changed as the paddle's orientation was varied. At the two ends of the orientation range, the laser emitted stable CW, whose polarization is linear and orthogonal to each other, indicating that they are the two stable principal polarization states of the laser emission. We then fixed the orientation of the paddle at a position within the stable square pulse emission range and further studied the features of the laser polarization switching with the pump strength change. At a weak pumping, stable square pulses could still be obtained. It was found that the antiphase intensity variation along the two orthogonal polarizations perfectly compensated each other. When the total laser emission intensity was measured, almost no signature of the polarization switching could be observed. However, as the laser emission intensity was increased, the antiphase polarization switching was no longer compensated. Fig 3a shows the oscilloscope trace of the total laser emission and one of the polarized emissions, respectively. Within one cavity roundtrip time there is one square-pulse along each polarization direction. Associated with one of the laser emission switching from one polarization to the other, an intensity dip appeared on the total laser emission. The profile of the intensity dip is stable with the cavity roundtrips, and each dip separates the two stable linear polarization states of the laser emission. Fig. 3b shows the corresponding optical spectra of the laser emissions. Laser emissions along the two orthogonal polarization directions have obvious different wavelengths and spectral distributions, showing that the coupling between the two polarization components is incoherent. In our experiments the cavity birefringence could be altered, by rotating the paddles of the PC or carefully bending the cavity fibers, eventually the wavelength separation between the two spectral peaks could be changed. Independent of the wavelength difference the intensity dip could always be obtained. Moreover, the width and depth of the dip varied with both the cavity birefringence and the pumping strength. At even higher pump strength, splitting of the square pulse could occur. Within one cavity roundtrip another square pulse could suddenly appear. The new square pulse was found unstable. It slowly moved in the cavity and eventually merged with the old one.

The intensity dips possess the characteristics of vector dark PDW soliton predicted by Haelterman and Shepperd [9], despite of the fact that the two stable polarization domains are now orthogonal linear polarizations instead of circular polarizations. To confirm that such PDWs could exist in our laser, we further numerically simulated the operation of the laser, using the model as described in [14] but with no polarizer in cavity. To make the simulation possibly close to the experimental situation, we used the following parameters: $\gamma$=3 W$^{-1}$km$^{-1}$, $\Omega_g$ =16 nm, $k''_{SMF}$= -23 ps$^2$/km, $k''_{EDF}$= -13 ps$^2$/km, $k'''$= -0.13 ps$^3$/km, $E_{sat}$=10 pJ, cavity length L= 11.4 m, $L_b$=L and G=120 km$^{-1}$. To favor the creation of an incoherently coupled domain wall, a perfect polarization switching inside the simulation window was put in the initial condition. This corresponds to the initial existence of the linear polarization switching in our laser caused by the laser gain competition and cavity feedback.

A stable PDW soliton separating the two principal linear polarization states of the cavity could be numerically obtained, as shown in Fig. 4. Fig. 4a shows the domain walls along each of the polarizations and Fig. 4b is the vector dark PDW soliton formed on the total laser emission and the ellipticity degree of the soliton. We adopted the definition of ellipticity degree q= (μ-ν)/ (μ+ν), where q=±1 represents the two orthogonal linearly polarized states and q=0 refers to a circularly polarized state [9]. The PDWs and soliton are stable and invariant with the cavity roundtrips. Numerically it was observed that even with very weak cavity birefringence, e.g. $L_b$=100L, stable PDW soliton could still be obtained. However, if the cavity birefringence becomes too large, e.g. larger than $L_b$=0.5L, no stable PDWs could be obtained.

Therefore, based on the numerical simulation and the features of the experimental phenomenon, we interpret the intensity dips shown in Fig. 3a as a type of vector dark PDW soliton. To understand why the PDWs and vector dark soliton could be formed in our laser, we note that Malomed had once theoretically studied the interaction of two orthogonal linear polarizations in the twisted nonlinear fiber [15]. It was shown that PDWs between the two orthogonal linear polarizations of the fiber exist, and the fiber twist could give rise to an effective force driving the domain walls. Considering that both the gain competition and the cavity feedback could have the same role as the fiber twist, not only the domain walls but also the moving of the domain walls could be explained.

In conclusion, we have reported the experimental observation of PDWs and vector dark PDW soliton in a linear birefringence cavity fiber ring laser. The domain walls and solitons are found to separate the two stable orthogonal linear principal polarizations. We have further shown that the cavity feedback and the gain competition could have played an important role on the formation of such PDWs and the vector dark domain wall soliton.

**Figure captions:**

Fig. 1: Schematic of the experimental setup. EDF: Erbium doped fiber. WDM: wavelength division multiplexer. DCF: dispersion compensation fiber. PC: polarization controller.

Fig. 2: Duration variation of the square pulses versus the orientation angle of one of the paddles of the intra-cavity PC.

Fig. 3: Vector dark polarization domain wall soliton emission of the laser. (a) Total laser emission (upper trace) and one of the polarized laser emissions (lower trace). (b) The corresponding optical spectra.

Fig. 4: Polarization domain wall numerically calculated. (a) Domain wall profiles. (c) The vector domain wall soliton and its ellipticity degree.

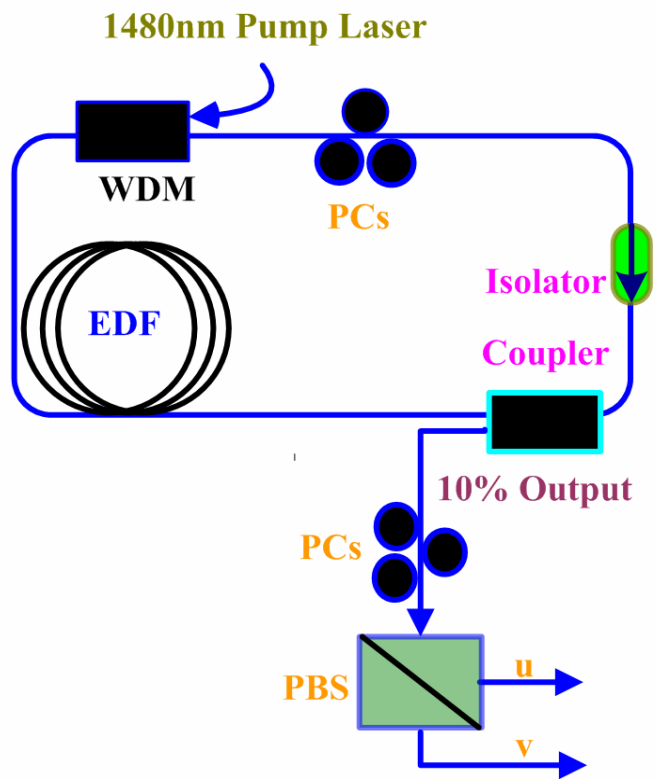

Fig. 1　H. Zhang et al.

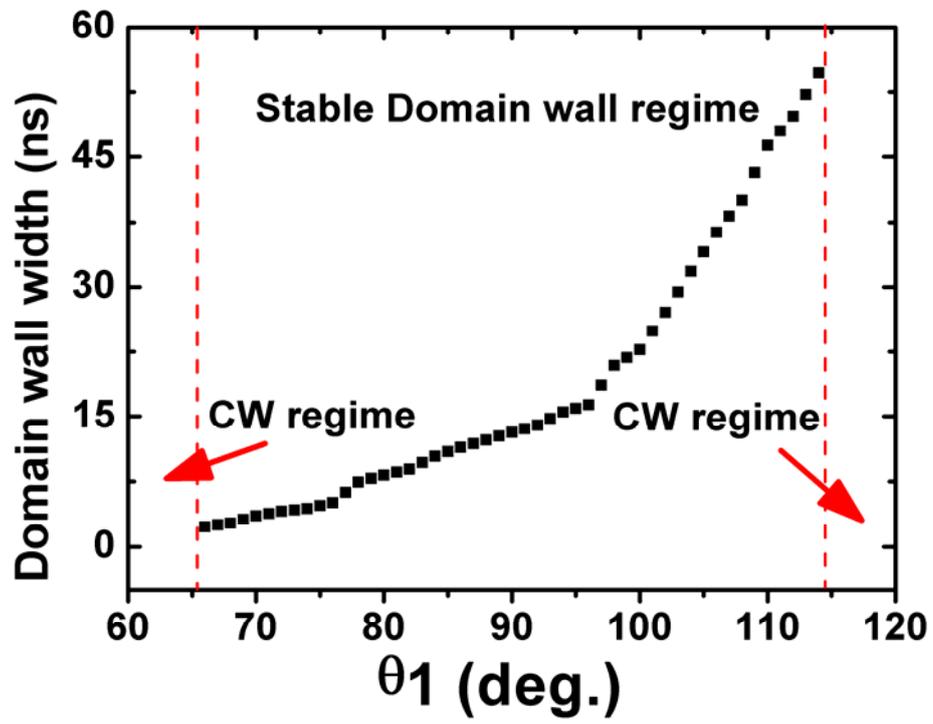

Fig. 2  H. Zhang et al.

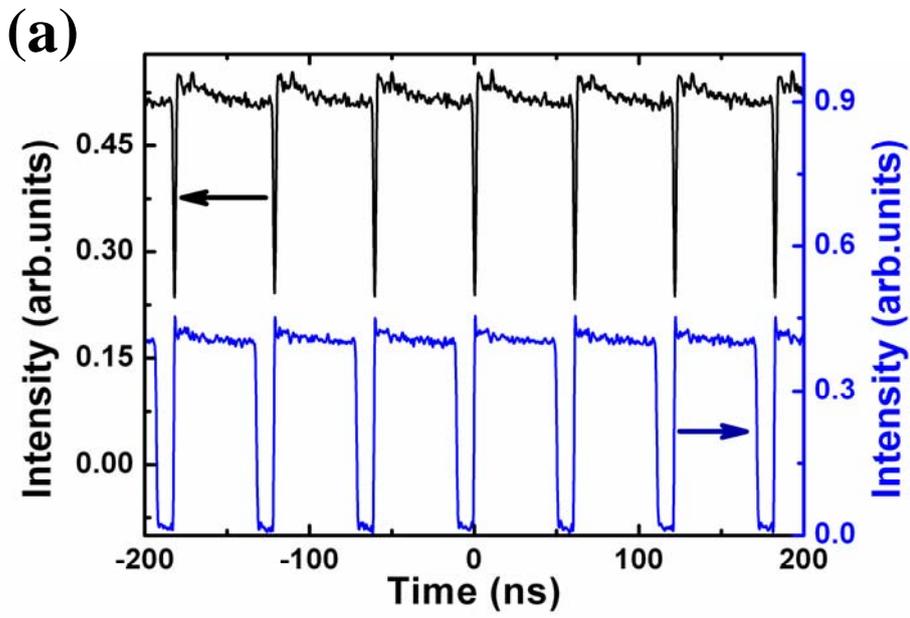

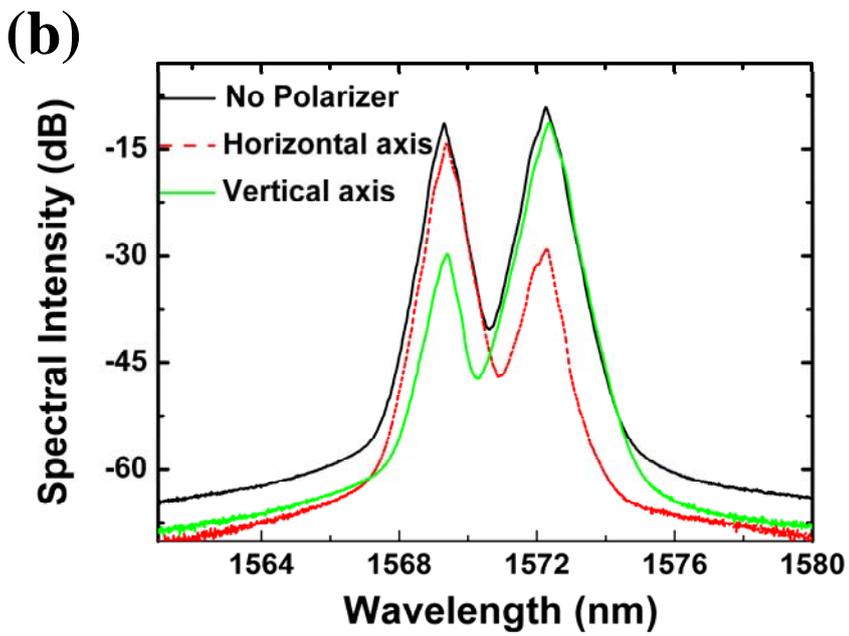

Fig. 3  H. Zhang et al.

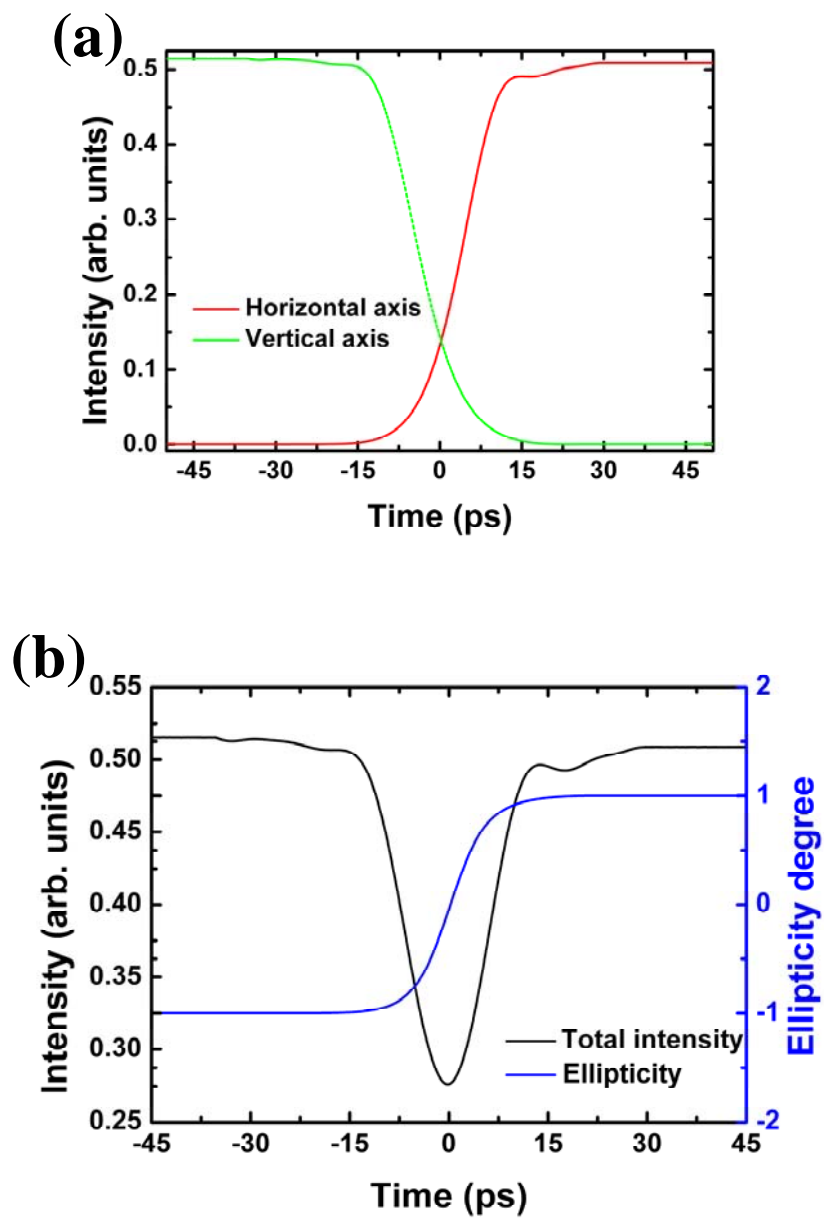

Fig. 4    H. Zhang et al.